\pdfoutput=1
\documentclass[%
 reprint,
 amsmath,amssymb,
 aps,
]{revtex4-1}

\usepackage{graphicx}
\usepackage{dcolumn}
\usepackage{bm}

\begin{document}


\title{A Scalable Algorithm to Explore the Gibbs Energy Landscape of Genome-scale Metabolic Networks}

\author{
Daniele De Martino$^{1}$, 
Matteo Figliuzzi$^{1}$,
Andrea De Martino$^{1,2,\star}$, 
Enzo Marinari$^{1,\star}$
}
\affiliation{%
$^1$ Dipartimento di Fisica, Sapienza Universit\`a di Roma, p.le A. Moro 2, 00185 Roma (Italy)\\
$^2$ IPCF-CNR, Unita\`a di Roma-Sapienza, Roma (Italy)\\
$^\star$These authors contributed equally to this work
}%


\begin{abstract}
{\bf Abstract -- }The integration of various types of genomic data into predictive models of biological networks is one of the main challenges currently faced by computational biology. Constraint-based models in particular play a key role in the attempt to obtain a quantitative understanding of cellular metabolism at genome scale. In essence, their goal is to frame the metabolic capabilities of an organism based on minimal assumptions that describe the steady states of the underlying reaction network via suitable stoichiometric constraints, specifically {\it mass balance} and {\it energy balance} (i.e. thermodynamic feasibility). The implementation of these requirements to generate viable configurations of reaction fluxes and/or to test given flux profiles for thermodynamic feasibility can however prove to be computationally intensive. We propose here a fast and scalable stoichiometry-based method to explore the Gibbs energy landscape of a biochemical network at steady state. The method is applied to the problem of reconstructing the Gibbs energy landscape underlying metabolic activity in the human red blood cell, and to that of identifying and removing thermodynamically infeasible reaction cycles in the {\it Escherichia coli} metabolic network (iAF1260). In the former case, we produce consistent predictions for chemical potentials (or log-concentrations) of intracellular metabolites; in the latter, we identify a restricted set of loops (23 in total) in the periplasmic and cytoplasmic core as the origin of thermodynamic infeasibility in a large sample ($10^6$) of flux configurations generated randomly and compatibly with the prior information available on reaction reversibility. \\ ~\\ {\bf Author Summary -- }The operation of biological systems is constrained under all circumstances by the laws of physics. Thermodynamics, in particular, dictates preferential directions in which biochemical reactions should flow at stationarity. When applied to cellular reaction systems (like metabolic networks), it favors the emergence of some (thermodynamically feasible) ways to organize the flow of matter while prohibiting others. The development of detailed predictive models for the biochemical activity of a cell relies on the possibility to integrate the laws of thermodynamics in genome-scale reconstructions of cellular metabolic networks. In this work we have devised an efficient relaxation algorithm to implement thermodynamic constraints in genome-scale models. Besides allowing to check for thermodynamic feasibility of reaction flow configurations, it is also capable of providing information on other relevant physico-chemical quantities. We have applied it to two cellular metabolic networks of different complexity, namely that of human red blood cells and that of the bacterium {\it Escherichia coli}. In the former case, we have obtained predictions for the intracellular chemical state (in terms of metabolite concentrations and reaction free energies) consistent with empirical knowledge; in the latter, we have effectively corrected thermodynamically infeasible flux configurations. 
\end{abstract}

\pacs{Valid PACS appear here}
\maketitle


\section*{Introduction}

Constraint-based models of cellular metabolism are important tools to analyze and predict the chemical activity and response to perturbations of cells without relying on kinetic details that are often unavailable. In such frameworks, the metabolic capabilities of a cell are inferred from the overall configuration space compatible with minimal physico-chemical constraints describing the non-equilibrium steady state of the underlying reaction network. First, feasible reaction flux vectors need to satisfy mass-balance conditions. Then, according to the second law of thermodynamics, in an open chemical network at steady state and constant temperature and pressure the direction of each reaction should ensure a decrease in Gibbs energy. Thermodynamic consistency of flux configurations satisfying mass-balance alone is in general not guaranteed due to the presence of infeasible cycles \cite{book,beard1,beard2}, even if reaction reversibility is pre-assigned based on careful estimations of chemical potentials in physiologic conditions \cite{alberty} (a procedure that was recently extended to genome-scale \cite{fleming,hatz}). Besides the flux organization, several other aspects involved in the analysis of genome-scale metabolic networks hinge directly on the explicit inclusion of thermodynamic constraints into the models, like the estimation of metabolite concentrations or the identification of reactions subject to regulation \cite{kummel2}. 

Much work has been concerned with implementing thermodynamic constraints in genome-scale models of metabolism. The removal of thermodynamic inconsistencies was proved to be useful in estimating concentrations and affinities besides fluxes in Flux-Balance-Analysis \cite{kau,palrev}, whose goal is to identify mass-balanced flux configurations maximizing a pre-determined physiological objective function \cite{sauer,biomass}. This has been achieved for instance by building mixed integer-linear or non-linear optimization problems that minimize the Euclidean distance of concentration levels from experimentally known values \cite{hoppe}, or ensure the absence of cycles in the resulting flux pattern \cite{tbmfa,cobrall,hepat}. On the other hand, information on feasible Gibbs energy ranges can be retrieved by exploiting the patterns of reaction interconnections encoded in the stoichiometry to narrow the experimental bounds \cite{kummel2}. These procedures may however require reliable prior thermodynamic information on metabolites and/or reactions, a type of knowledge that is often unavailable.  An important lesson of this kind of approaches is that the key input for the thermodynamic profiling of a reaction network is often provided by the stoichiometric matrix \cite{beardstoc}.

The scalability of algorithms to solve mixed integer-linear (or non-linear) programming problems may become an issue when the underlying network size is large or when one is interested in sampling the solution space (for both free energies and fluxes) rather than focusing on a potentially small set of configurations (e.g. optima). Luckily, however, solutions to computationally hard problems can often be generated efficiently with the help of heuristic algorithms based on simple local rules. The use of message-passing algorithms to characterize the high-dimensional volume of the solution space of FBA models \cite{BMC} (with a convex, continuous solution space) or to solve large combinatorial constraint-satisfaction problems \cite{KSAT} (with a discrete and possibly fragmented solution space) is an example of the success of this kind of strategy.

Our goal in this paper is to obtain information about the landscape of Gibbs free energies compatible with a given vector of reaction directions by following a route that allows to use all stoichiometric information via heuristics inspired by perceptron learning. In a nutshell, the method we propose consists in exploiting the network's structure to iteratively build up correlations between the chemical potentials of the reacting species starting from a seed of empirical biochemical knowledge, until a thermodynamically consistent profile is achieved. The resulting algorithm is completely scalable and can be employed for different purposes, like checking the feasibility of flux configurations, identifying and removing infeasible cycles, estimating reaction affinities, and obtaining bounds for (log-)concentrations and free energies of formation. In the following, we describe the method in detail, providing a mathematical proof of convergence as well as theoretical arguments highlighting the main idea behind the procedure. As applications, we focus on two metabolic networks of rather different complexity. First, we shall obtain a detailed reconstruction of the Gibbs energy landscape underlying metabolic activity in the human red blood cell (hRBC) starting from the flux maps obtained in \cite{palsson,andrea}. Then, the metabolic network of {\it {\it Escherichia coli}}, iAF1260 \cite{bigcoli}, will be analyzed to eliminate infeasible cycles from randomly generated flux configurations.

\section*{Materials and Methods}

\subsection*{Materials}

The cellular systems analyzed in this study are (i) the model of the hRBC metabolism developed in\cite{Wiback:2002fk} and discussed in \cite{palsson}, and (ii) the reconstructed metabolic network of the bacterium {\it {\it Escherichia coli}} iAF1260, presented in \cite{bigcoli}. The former consists of 35 intracellular reactions among $39$ metabolites subject to $12$ uptake fluxes. The latter includes $2381$ reactions among $1039$ metabolites. The basic information extracted from these models is the $M\times N$ stoichiometric matrix (with $M$ and $N$ the number of metabolites and reactions, respectively), denoted below as $\mathbf{S}$. For {\it Escherichia coli}, we will consider the inner matrix of periplasmic and cytoplasmic reactions without repetitions, which consists of  $1767$ reactions among $1349$ chemical species, once periplasmic and cytoplasmic metabolites are distinguished. According to the reversibility assignment given in \cite{bigcoli}, $1475$ of the $1767$ processes above are unidirectional, with $292$ being reversible. The biochemical data we shall refer to or make use of include the standard free energies of formation of metabolites, given in \cite{kummel}, where they are computed at $T=298 \mathrm{~K}$ , $P=1\mathrm{~atm}$, $\mathrm{pH} =7.6$ and ionic strenght $0.15\mathrm{~M}$, according to the prescriptions of \cite{alberty}.  The estimated intracellular concentration ranges for the hRBC were extracted from the Bionumbers database \cite{Milo:2010fk} and refer to measurements in different settings. It is worth to notice that the experimental errors on such values reflect the intrinsic uncertainty due to statistical cell-to-cell fluctuations (since measurements of concentration levels are usually carried out by averaging over numbers of cells ranging from $10^2$ to $10^8$) and analytical error. The stoichiometric matrix and thermodynamic potentials employed for the analysis of the hRBC are respectively available as Supporting Datasets S1 and S2.

\subsection*{Methods}

\subsubsection*{Algorithm to compute chemical potentials}

According to the second law of thermodynamics, in an open system at constant temperature $T$ and pressure $P$ the Gibbs energy $G=E-PV-TS$ (where $E$, $V$, $S$ are respectively the energy, volume and entropy of the system) never increases spontaneously. This means that the direction $u_i\in\{-1,1\}$ ($+1$ for forward, $-1$ for backward) of every chemical reaction $i\in\{1,\ldots, N\}$ occurring in the system should be opposite to the Gibbs energy change $\Delta G_i$ induced by reaction $i$, i.e. 
\begin{equation}\label{tc}
u_i \Delta G_i \leq 0~~~~~\forall i\;.
\end{equation} 
The equality holds if reaction $i$ is in equilibrium. Denoting by $S_{\alpha,i}$ the stoichiometric coefficient of reactant $\alpha\in\{1,\ldots,M\}$ in reaction $i$, with the standard sign convention to distinguish substrates  ($S_{\alpha,i}<0$) from products ($S_{\alpha,i}>0$), the vector of $\Delta G_i$'s for a well-mixed system can be written in terms of the chemical potentials $\boldsymbol{\mu}=\{\mu_\alpha\}$ (where $\mu_\alpha$ is the Gibbs energy per mole of species $\alpha\in\{1,\ldots,M\}$) as (see e.g. \cite{book}, Ch. 1)
\begin{equation}\label{four}
\boldsymbol{\Delta} \mathbf{G} = \mathbf{S}^T \boldsymbol{\mu}
\end{equation} 
Given flux vectors $\mathbf{v}$ such that $\mathbf{S} \mathbf{v}=0$ (i.e. steady-state flux configurations), equation (\ref{four}) implies that $\mathbf{v}\cdot\boldsymbol{\Delta} \mathbf{G}=0$, i.e. that the `loop law' holds. The Gibbs energy landscape reconstruction problem consists, given a vector $\mathbf{u}=\{u_i\}$ of reaction directions ($u_i\in\{-1,1\}$), in generating vectors $\boldsymbol{\mu}$ that satisfy the system of linear inequalities
\begin{equation}\label{constr}
x_i\equiv -u_i  \sum_{\alpha=1}^M S_{\alpha,i} \mu_\alpha \geq 0~~~~~\forall i\;.
\end{equation}
Note that the solution space of (\ref{constr}) for fixed directions is convex (while non-convexity can arise if directions are allowed to vary). Relaxation methods (see e.g. \cite{LP}, Ch. 12, or \cite{gof}) are among the most effective procedures to find solutions of systems such as (\ref{constr}). These techniques date back at least to the Jacobi method for solving systems of linear equalities, and were extended to inequalities in the 1950's \cite{agmon,motz}. In essence, they are iterative methods in which variables are updated so that at every iteration one of the violated inequalities is fixed. While this readjusts the entire vector without a guarantee that constraints that were previously satisfied will be broken, convergence to a solution (if it exists) is guaranteed if the update step is chosen wisely. We shall employ a relaxation algorithm known as MinOver, which was developed in the context of neural network learning \cite{mez}, and has been employed, in a slightly modified form \cite{jstat}, to explore the space of flux states compatible with minimal stability constraints {\it \`a la Von Neumann} \cite{pnas,kyoto}. Figure 1 displays a flowchart of the procedure for the present case. 
\begin{figure}
\begin{center}
\includegraphics[width=8.5cm]{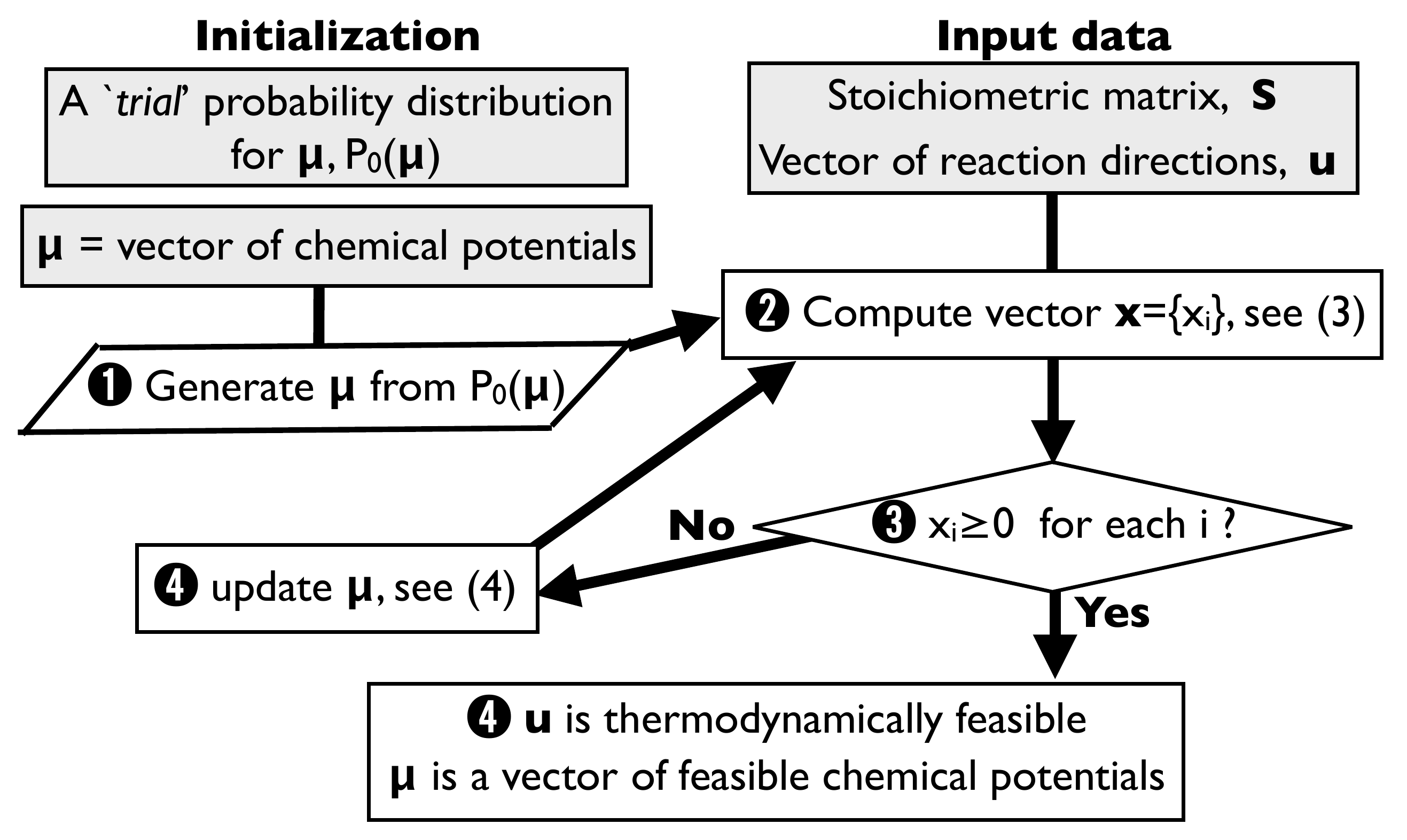}
\caption{\label{uno} {\bf Flowchart of the algorithm.} Given a stoichiometric matrix $\mathbf{S}$ and a generic vector $\mathbf{u}$ of reaction directions, the algorithm generates a vector $\boldsymbol{\mu}$ of chemical potentials if $\mathbf{u}$ is thermodynamically feasible. $\mathbf{u}$ may for instance be taken from a steady state flux configuration.}
\end{center}
\end{figure}
One starts from a `trial' probability distribution $P_0(\boldsymbol{\mu})$ of chemical potential vectors. Its role for the moment is simply that of initializing the algorithm, which is done by generating a random vector $\boldsymbol{\mu}$ under $P_0(\boldsymbol{\mu})$. For simplicity, one may think that $P_0(\boldsymbol{\mu})=\prod_{\alpha=1}^M P_0^\alpha(\mu_\alpha)$, with prescribed distributions $P_0^\alpha$, e.g. uniform over a given interval: in this case each initial $\mu_\alpha$ is selected randomly and independently from its trial distribution $P_0^\alpha$. On the other hand, $P_0^\alpha$ might contain prior biochemical information, e.g. by being a uniform distribution centered around the known experimental values of $\mu_\alpha$ and of sufficiently large width to span several orders of magnitude in concentrations. (The precise construction of $P_0(\boldsymbol{\mu})$ for our case studies is discussed below.) The algorithm is based on the following steps: 
\begin{enumerate}
\item Generate a chemical potential vector $\boldsymbol{\mu}=\{\mu_\alpha\}$ from
  $P_0(\boldsymbol{\mu})$.
\item Compute $\mathbf{x}=\{x_i\}$ from (\ref{constr}) and
  $i_0=\text{arg }\min_i x_i$ (i.e., $i_0$ is the index of the least satisfied constraint).
\item If $x_{i_0}\geq 0$ then $\boldsymbol{\mu}$ is a thermodynamically
  consistent chemical potential vector for $\mathbf{u}$; exit (or go
  to 1 to obtain a different solution).
\item If $x_{i_0}<0$, update $\boldsymbol{\mu}$ as 
\begin{equation}\label{deltag}
\boldsymbol{\mu}~\to~\boldsymbol{\mu}-\lambda u_{i_0} \mathbf{S}_{i_0}
\end{equation}
(where $\lambda>0$ is a constant and $\mathbf{S}_{j}$ is the $j$-th column of matrix $\mathbf{S}$), go to 2 and iterate.
\end{enumerate}
As is generally true in MinOver schemes, the reinforcement term in
(\ref{deltag}) drives the gradual adjustment of chemical potentials by ensuring
that, at every iteration, the least satisfied constraint (labeled
$i_0$) is improved. Convergence to a solution, if one exists, is guaranteed for any $\alpha>0$. To see it, suppose that a vector $\boldsymbol{\mu}^\star$ exists, such that
\begin{equation}
 -u_i (\mathbf{S}_i \cdot \boldsymbol{\mu}^\star) \geq c~~~~~\forall i\;,
\end{equation} 
with $c>0$ a constant (in other words, $\boldsymbol{\mu}^\star$ is a solution of (\ref{constr})). From Eq.  (\ref{deltag}), the chemical potential vector at the $\ell$-th iteration step, $\boldsymbol{\mu}(\ell)$, satisfies
\begin{eqnarray}
\boldsymbol{\mu}(\ell) \cdot \boldsymbol{\mu}^\star &=& \boldsymbol{\mu}(\ell-1) \cdot \boldsymbol{\mu}^\star - \lambda s_{i_0(\ell-1)} (\mathbf{S}_{i_0(\ell-1)} \cdot \boldsymbol{\mu}^\star ) \nonumber \\
& \geq &  \boldsymbol{\mu}(\ell-1) \cdot \boldsymbol{\mu}^\star + \lambda c  \nonumber \\
 & \geq & \boldsymbol{\mu}(0) \cdot \boldsymbol{\mu}^\star + \ell \lambda c \;,
\end{eqnarray}
where $i_0(\ell)$ is the value taken by $i_0$ at step $\ell$. Similarly, one finds 
\begin{equation}
\boldsymbol{\mu}(\ell) \cdot \boldsymbol{\mu}(\ell) \leq \boldsymbol{\mu}(0)\cdot\boldsymbol{\mu}(0) + \ell \lambda^2 A\;,
\end{equation}
with $A = \max_i \sum_\alpha (S_{\alpha,i})^2$. As a consequence,
\begin{equation}
d(\ell)\equiv \frac{\boldsymbol{\mu}(\ell)\cdot \boldsymbol{\mu}^\star}{|\boldsymbol{\mu}(\ell)| |\boldsymbol{\mu}^\star|} \geq \frac{ \boldsymbol{\mu}(0) \cdot \boldsymbol{\mu}^\star + \ell \lambda c }{ |\boldsymbol{\mu}^\star| \sqrt{ |\boldsymbol{\mu}(0)|^2 + \ell \lambda^2 A }} 
\end{equation}
However by the Cauchy-Swartz inequality $d(\ell) \leq 1$, and an upper bound for the number of steps can be obtained from $d(\ell_c)=1$ which to leading orders in $1/c$ and $1/\lambda$ gives
\begin{equation}
\ell_c\simeq \frac{\delta_1}{c^2}+\frac{\delta_2}{\lambda^2}+\frac{\delta_3}{\lambda c}~,
\end{equation}
where $\delta_1>0$, $\delta_2>0$ and $\delta_3$ are constants proportional to $A$. Hence starting from an initial random vector of chemical potentials the algorithm is able to obtain a new vector ensuring that (\ref{constr}) is satisfied. Re-initializing the algorithm from a different random vector sampled from $P_0(\boldsymbol{\mu})$ allows to retrieve another solution and in turn explore the space of $\boldsymbol{\mu}$'s satisfying (\ref{constr}). 

In essence, the final outcome of multiple (random) initializations of the above algorithm is a set of {\it correlated} probability distributions for the $\mu_\alpha$'s (at odds with the $P_0(\boldsymbol{\mu})$, which was assumed to be a product measure, so that no correlations were present initially).  The algorithmic origin of such interdependencies can be understood
considering that chemical potentials are updated dynamically through a
series of reinforcement steps of the form $\lambda u_i S_{\alpha,i}$. It
follows that the final value of $\mu_\alpha$ can be written as
$\mu_\alpha=\mu_\alpha^{\text{tr}}+\alpha\sum_i h_i S_{\alpha,i}$, where
$\mu_\alpha^{\text{tr}}$ is the trial chemical potential sampled from $P_0$ (i.e. the initial value of $\mu_\alpha$) and $h_i\in\{0,1\}$ is an index which is updated (increased or decreased by one
according to the sign of the reaction) each time reaction $i$ tries to
invert. The connected correlations $\langle \mu_\alpha \mu_\beta \rangle_c\equiv  \langle \mu_\alpha \mu_\beta\rangle - \langle \mu_\alpha \rangle \langle \mu_\beta \rangle$ between chemical potentials  (where $\langle\cdots\rangle$ is an average over all possible choices of the initial conditions) can thus be decomposed as
\begin{multline}\label{correlations}
\langle \mu_\alpha \mu_\beta \rangle_c=\delta_{\alpha\beta}\sigma^2_\alpha+\lambda\sum_{i=1}^N 
S_{\alpha,i}\langle \mu_\beta^{\text{tr}}  h_i\rangle_c+\\+\lambda\sum_{i=1}^N S_{\beta,i}
\langle \mu_\alpha^{\text{tr}} h_i\rangle_c+\lambda^2\sum_{i,j}S_{\alpha,i} 
S_{\beta,j}\langle h_i h_j \rangle_c\;,
\end{multline}
where $\sigma_\alpha^2$ is the variance of $P_0^\alpha$ and $\delta_{\alpha\beta}=1$ if $\alpha=\beta$ and $=0$ otherwise (so that $\langle \mu_\alpha^{\text{tr}} \mu_\beta^{\text{tr}} \rangle_c=\delta_{\alpha\beta}\sigma^2_\alpha$). 
In the Gaussian approximation for $\sum_{i=1}^N S_{\alpha,i} h_i$, the leading term (of $\mathcal{O}(N)$) in the above sum is
\begin{equation}\label{appro_correlations}
\lambda^2\sum_{i=1}^NS_{\alpha,i} S_{\beta,i} \sigma^2_{h_i}~~~~~(\alpha\neq\beta)\;.
\end{equation}
Therefore to leading order, the dynamics tends to correlate (resp. anti-correlate) the chemical potentials of metabolites typically appearing on the same (resp. opposite) side of the reaction equations. In a sense, the above scheme allows to modify $P_0$ by building up correlations between chemical potentials according to the interconnections encoded in $\mathbf{S}$. Note that, at odds with the method proposed in \cite{kummel2}, the resulting $\mu_\alpha$'s can exceed the initial bounds defined by $P_0(\boldsymbol{\mu})$.

If there is no prior information on the direction of some reactions (e.g. because they are putatively reversible), the corresponding constraints (\ref{constr}) are formally absent, as if $u_i=0$. However, the above method still allows to retrieve information about the chemical potential of a metabolite involved in them, provided it is not employed in reversible processes only, in which case its $\mu_\alpha$ clearly is never updated. 

Finally, some observations are in order about the solution space of (\ref{constr}), which in general has the form of an unbounded cone passing trough the origin. If one is interested in uniform sampling the space of $\boldsymbol{\mu}$'s making $\mathbf{S}$ thermodynamically feasible, boundedness is an essential precondition. The simplest way to obtain a bounded solution space consist in clamping some $\mu_\alpha$'s, i.e. in keeping them fixed at definite values  throughout the updating. Note that fixing some $\mu_\alpha$'s is also crucial to set a scale for chemical potentials. The same effect can be achieved by assigning hard ranges of variability for potentials in the form of bounds like $\mu_\alpha^{\text{min}}\leq \mu_\alpha\leq \mu_\alpha^{\text{max}}$  (e.g. according to experimental or biochemical information). Such inequalities can simply be added to the list of thermodynamic constraints (\ref{constr}). Alternatively, one may add other types of global, non-homogeneous constraints bearing a physical justification. For instance, if uptake fluxes are included in $\mathbf{S}$, the chemical potentials of the external metabolites should be fixed; or, volume constraints for the feasible ranges of concentrations could be added if the standard free energies of formation are known. Once this is done, the system (\ref{constr}) becomes inhomogeneous and its solution space is formed by the union of a convex polyhedron and a cone; boundedness can be achieved if and only if the cone shrinks to the null vector, which occurs if the related homogeneous system of equations has no solutions apart from the trivial one, $\mu_\alpha=0~\forall\mu$ \cite{solodov}. In synthesis, one can obtain a bounded solution space for (\ref{constr}) by clamping the chemical potential of a sufficient number of metabolites to ensure that the homogeneous system of equalities associated to the inhomogeneous system of inequalities obtained by clamping some $\mu_\alpha$'s in (\ref{constr}) admits the null vector as its only solution.

 A number of interesting theoretical and computational questions arise at this stage, regarding e.g. the minimal amount of prior information on chemical potentials needed to bound the solution space of (\ref{constr}), or computationally efficient and scalable methods to obtain uniform sampling (besides Monte Carlo, which may be infeasible at high dimensions as suggested by the ``curse of dimensionality'', see e.g. \cite{Schellenberger:2009fk}). To our knowledge, there is no mathematical proof that MinOver schemes are capable of sampling a bounded solution space uniformly, although low dimensional tests suggest that this might indeed be the case \cite{pnas}. Our goal in the present paper, rather than uniformly sampling the space of chemical potentials granting feasibility, is that of exploring feasible configurations ``close'' to the prior biochemical information we have injected. To quantify this idea, we have explicitly compared the solutions obtained via the above procedure with those retrieved from the minimization of a cost function (the average Euclidean distance $\overline{d}$ between the prior and the solution) and by a standard relaxation method (see Results: Exploring the Gibbs energy landscape of the hRBC metabolic network). The heuristics we present indeed turns out to roughly minimize $\overline{d}$, with the advantage of being considerably faster than a penalty method. In addition, it allows to access refined information on the Gibbs energy landscape (i.e. compute chemical potentials and Gibbs energy changes) even when the initialization is complemented by a noisy or inconsistent biochemical prior, since the algorithm correctly identifies and gradually removes inconsistencies. The solutions thus obtained identify a restricted and statistically well-behaved set of chemical potentials with physiological significance.  We are currently unable to go beyond this point. In addition, we shall see that by the same method one can verify the thermodynamic consistency of, and eventually adjust, specific flux configurations of biochemical reaction networks. 

\subsubsection*{Algorithm to identify and remove loops}

The algorithm just discussed generates chemical potential vectors given a thermodynamically feasible vector of reaction directions. A generic assignment of reaction directions, however, could be such that the system (\ref{constr}) has no solutions apart from the trivial one. In accordance with the Farkas-Minkowski lemma \cite{solodov} this happens if and only if there is at least one infeasible loop, i.e. if there is a set $\mathcal{L}$ of intracellular reactions for which positive constants 
$k_i>0$ exist such that
\begin{equation}\label{ul}
\sum_{i \in \mathcal{L}} k_i u_i S_{\alpha,i} = 0~~~~~\forall \alpha\;.
\end{equation} 
In presence of a loop, relaxation methods (MinOver included) do not converge, just because the least satisfied constraint moves along the loop causing the iteration to cycle indefinitely. More precisely, in presence of a loop the MinOver dynamics becomes periodic or almost periodic. For a periodic dynamics $\boldsymbol{\mu}(\ell+L)=\boldsymbol{\mu}(\ell)$ so that, using (\ref{deltag}), for any $\ell$ we have
\begin{equation}
0\equiv \mu_\alpha(\ell+L)-\mu_\alpha(\ell)=-\lambda \sum_{\tau=1}^L s_{i_0(\tau+\ell)}S_{\alpha,i_0(\tau+\ell)} \,\,\,\,\,\forall \alpha .\label{periodic}
\end{equation}
Comparing this with (\ref{ul}) it can be gathered that reactions updated over a period define a loop of length $L$ (setting $k_i$ to the number of times constraint $i$ has been updated). This suggests a simple way to correct configurations of reaction directions that are thermodynamically infeasible to start with:
\begin{enumerate}
\item[(i)] While running the algorithm to compute chemical potentials for a large number of iteration steps $T$, keep track of the last say $K$ least unsatisfied constraints, i.e. store $i_0(\ell)$ for $\ell=T-K+1,\ldots,T$, with  $K$ a reasonably large number (e.g. $500$), and count the number $n$ of different reactions appearing in the series.
\item[(ii)] Search, within such subset of reactions, for a loop of length $L$ by looking for solutions to equation (\ref{ul}) with $k_i\neq 0$ for $L$ reactions only, for all $L$-uples, starting from $L=3$ and increasing $L$. 
\item[(iii)] If a loop is found, change the direction of one of its reversible reactions chosen with uniform probability. 
\end{enumerate}
In the Results section we shall use this heuristics to spot loops and eliminate them in all of the infeasible configurations we shall generate for a large metabolic network (E. coli iAF 1260). For this network, it turns out that $n\leq 50$ in all runs and that accounting for short loops (of length up to $6$) suffices to correct all $10^5$ randomly-generated configurations we tested. This is rather important since, in principle, step (ii) of the above procedure could be exponential in $L$. 

A computer code implementing the algorithms to compute chemical potentials and identify and remove infeasible loops is downloadable from http://chimera.roma1.infn.it/SYSBIO/

\section*{Results}

\subsection*{Exploring the Gibbs energy landscape of the hRBC metabolic network}

As a first application, we have employed the MinOver scheme outlined above to analyze the thermodynamic landscape of the hRBC metabolic network. As a starting point, we have considered the flux configurations obtained in \cite{palsson} and \cite{andrea} respectively by Monte Carlo sampling of the solution space of mass balance equations (MBE) and by MinOver sampling of states compatible with  Von Neumann's constraints (VNC). In brief, MBE describe steady-state fluxes in terms of Kirchhoff-type laws enforced at metabolite nodes of the network as $\mathbf{Sv=0}$. In such a scenario, intracellular concentrations are clamped. VNC, instead, `soften' the mass-balance equations by requiring that, for intracellular metabolites, $\mathbf{Sv\geq 0}$. In the underlying steady state intracellular concentrations can grow in time if flux vectors allowing for it exist. Once the nutrient availability is set, VNC define a self-consistent flux problem where the system selects how much of the nutrients to use and, eventually, which metabolites are globally produced. For intracellular metabolites, VNC correspond to a stability requirement since, in a dynamical setup, a violation of one of the inequalities implies the existence of a metabolite whose total amount used as input in metabolic processes exceeds the total amount returned as output (see e.g. \cite{gale,Martino:2005mi}). VNC are also closely linked to the metabolite producibility problem introduced in \cite{ib00}. We shall thus make use of the reaction direction vectors $\mathbf{u}$ obtained in  \cite{palsson,andrea} for the hRBC. In summary:
\begin{itemize}
\item according to \cite{palsson}, the net flux of all reactions is in the forward direction, except PGI, R5PI and ApK, which are found to operate bidirectionally;
\item according to \cite{andrea}, the net flux of all reactions is in the forward direction, except R5PI (which is found to be operating bidirectionally), and PGI and ApK (which are found to have a net backward flux).
\end{itemize}
As a first step, we tested the thermodynamic feasibility of these direction assignments, 
solving (\ref{constr}) by starting from the vector $\mu^{\text{tr}}_\alpha=1$ $\forall \mu$. 
A solution is found for both MBE and VNC assignments. Both sets of assignments then turn out to be (expectedly) thermodynamically feasible. In this case, however, it is the emerging thermodynamic landscape that is of interest to us. As the initial distribution of chemical potentials $P_0$ we selected a product of independent uniform distributions $P_0^\alpha$. The $P_0^\alpha$ for compounds with known empirical bounds on concentrations were centered at a value $\langle \mu_\alpha\rangle$ computed from the Gibbs energy of formation $\mu_{0,\alpha}$ and the estimated intracellular concentration $c_\alpha$ under the hypothesis of a dilute solution, i.e. at
\begin{equation}\label{avg}
\langle \mu_\alpha\rangle = \mu_{0,\alpha} + RT \log{c_\alpha}\;,
\end{equation}  
and were taken to span two standard deviations in concentrations. The $P_0^\alpha$ for metabolites whose concentration estimates were unavailable (namely 6PGL, RL5P, X5P, R5P, S7P) were taken to be centered again at (\ref{avg}) but with $c_\alpha=10^{-4}\mathrm{~M}$, and were assumed to span four orders of magnitude uniformly in the chemical potential scale. Finally, the chemical potential of water was clamped.

Results for the estimated concentrations and Gibbs energy changes (computed from (\ref{four}) using the final chemical potential vectors) are showcased in Figures 2 and 3, respectively. 
\begin{figure}
\begin{center}
\includegraphics[width=8.5cm]{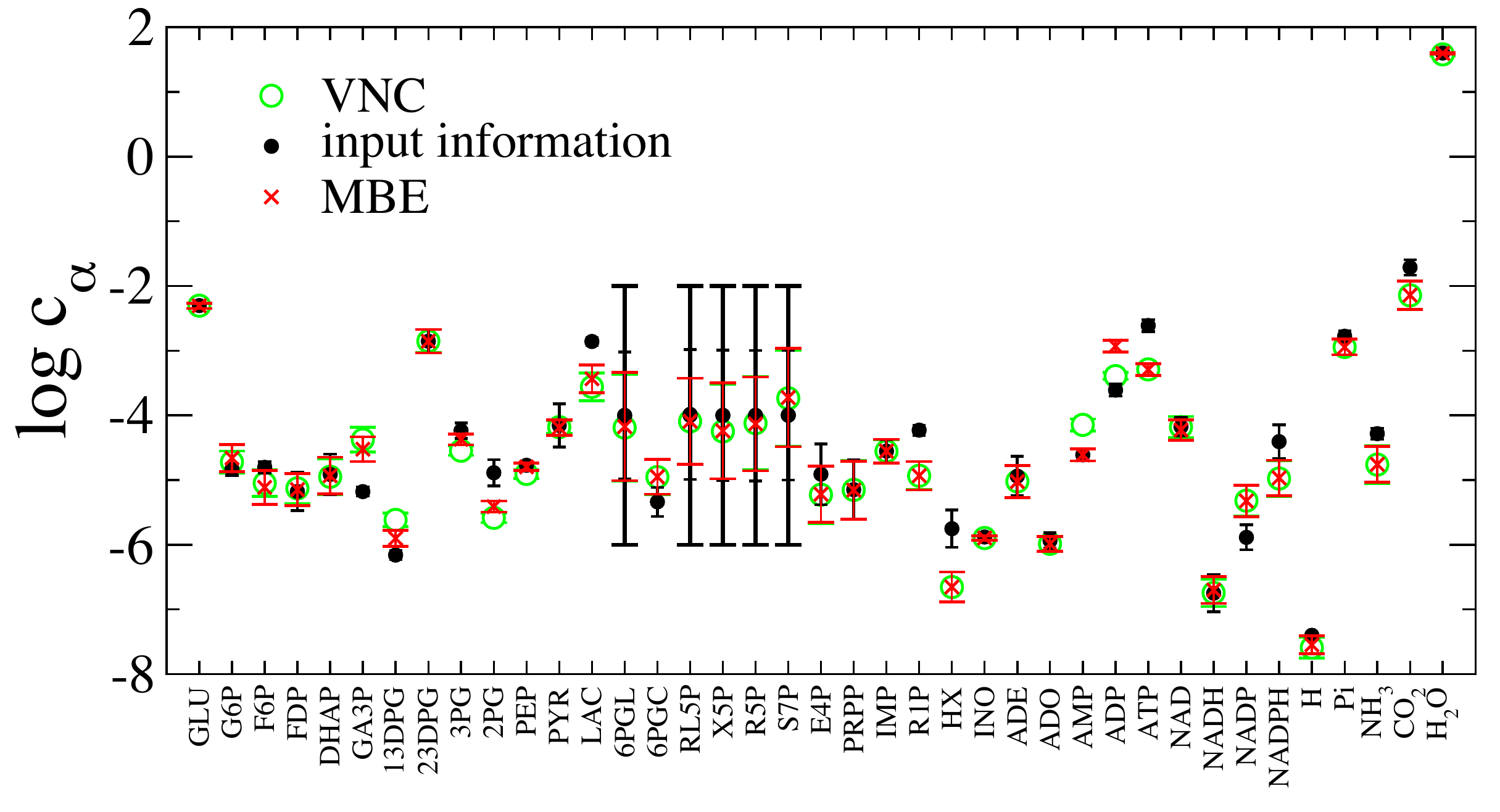}
\caption{{\bf Estimated log-concentrations of metabolites in the hRBC metabolic network.} The input information used to initialize the algorithm (with error bars) is denoted by black markers (see text for details). Values obtained starting from direction assignments corresponding to a sample of $10^5$ MBE and VNC solutions are shown respectively as red and green markers.}
\end{center}
\end{figure}
\begin{figure}
\begin{center}
\includegraphics[width=8.5cm]{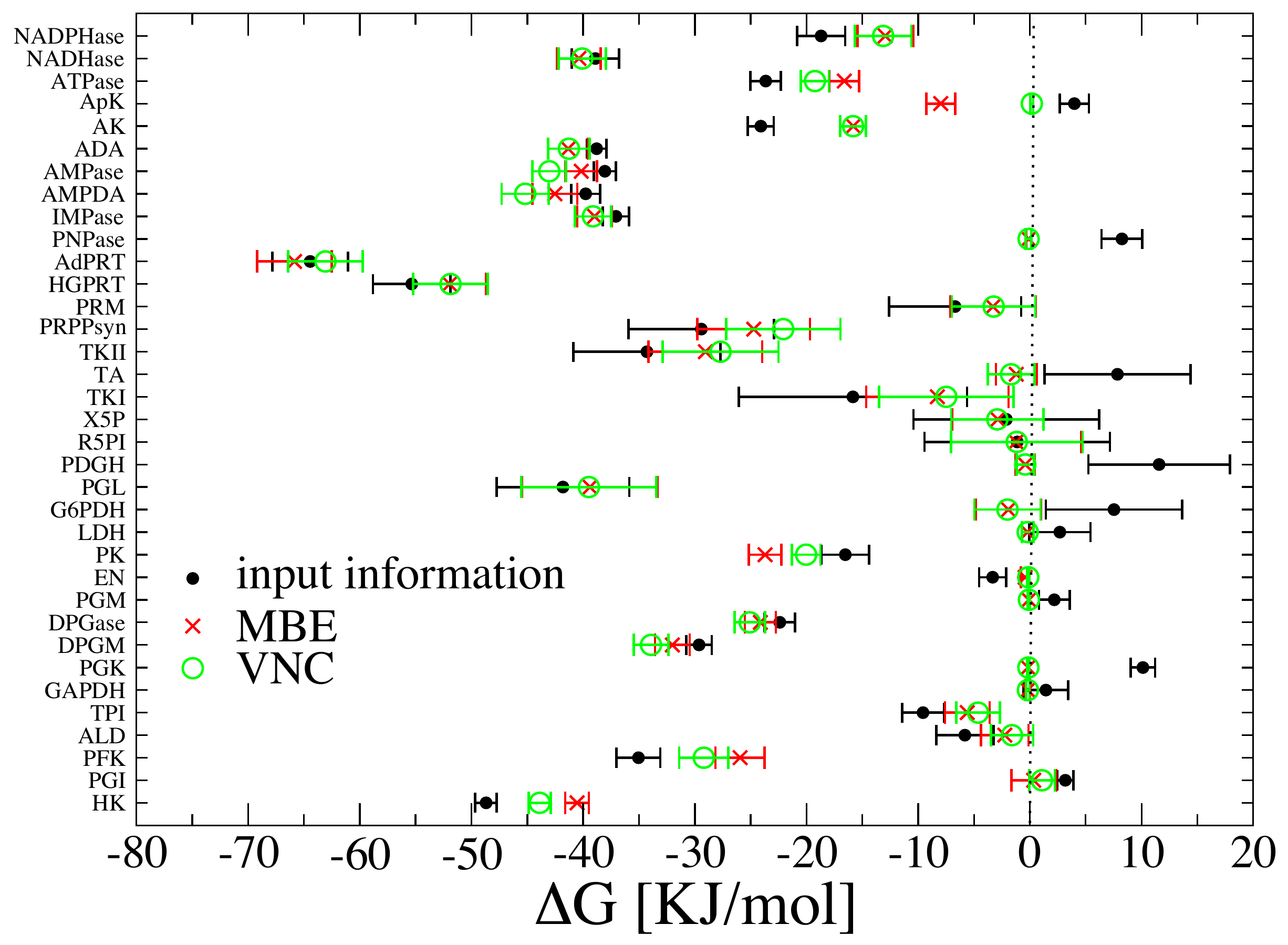}
\caption{{\bf Estimated Gibbs energy changes of reactions in the hRBC metabolic network.} The input information used to initialize the algorithm (with error bars) is denoted by black markers; the values obtained starting from direction assignments corresponding to MBE and VNC solutions are shown respectively as red and green markers. Note that the input information is consistent with reactions operating in the reverse direction for GAPDH, PGK, PGM, LDH, G6PDH, TA and PNPase. The algorithm is able to correct these inconsistencies starting from both MBE- and VNC-compatible direction assignments.}
\end{center}
\end{figure}
Note that several of the bounds encoded in $P_0$ (black markers) indicate that the Gibbs energy change in a reaction is positive and in contrast with the reaction direction assignment from the flux problem. Specifically this happens for GAPDH, PGK, PGM, LDH, G6PDH, TA and PNPase, due either to the actual experimental estimates for the affinities or to the initial uncertainty we place on concentrations. Such bounds turn out to be altered by MinOver in a direction compatible with direction assignments based on mass balance, as final affinities display significant changes with respect to the picture embedded in $P_0$. At the same time, we observe that in some cases the fluctuations of $\mu_\alpha$ do exceed the initial boxes defined by $P_0$, leading to an estimate for the concentration range also for metabolites whose level has not been experimentally probed (6PGL, RL5P, X5P, R5P, S7P). Our predictions for the levels of (1,3)-diphosphoglycerate ((1,3)-DPG), 2-phosphoglycerate (2PG) and phosphoenolpyruvate (PEP) slightly differ from the experimental estimates. This is most likely a consequence of the fact that we are forcing the phosphoglycerate kinase (PGK) and the glyceraldehyde phosphate dehydrogenase (GAPDH) reactions in the forward direction, in agreement with the steady state direction assignments for glycolysis, even if the experimental values would classify them as reversible. In addition, we obtain levels of key metabolites like ATP and inorganic phosphate that differ slightly from experimental estimates, while our predictions for ADP and AMP fail under the MBE and VNC direction assignments, respectively. On one hand this could be due to errors in the prior information on standard free energies but on the other hand, precise experimental estimates of the levels of such highly interchanging metabolites might be difficult to achieve.   

We can now quantify the extent to which the solutions we generate are ``close'' to the prior. In Figure 4 we compare the Gibbs energy changes obtained for the hRBC using MBE directions in three different ways: (a) the MinOver algorithm introduced here, with update given by (\ref{deltag}); (b) a penalty method defined by the update rule
\begin{equation}\label{penalty}
\mu_\alpha~\to~\mu_\alpha-\lambda_1\left[2(\mu_\alpha-\mu_\alpha(0))+\lambda_2\sum_{i\in\textsc{Unsat}} S_{\alpha,i}\right]~~~~~\forall\alpha
\end{equation} 
where $\lambda_1$ and $\lambda_2$ are constants and the sum extends over all $i$'s such that $x_i<0$; (c) a standard relaxation method defined by the update rule
\begin{equation}\label{relax}
\mu_\alpha~\to~\mu_\alpha-\frac{2x_{i_0}}{u_{i_0}\sum_\beta (S_{\beta,i_0})^2}~S_{\alpha,i_0}~~~~~\forall\alpha~~.
\end{equation} 
\begin{figure}
\begin{center}
\includegraphics[width=8.5cm]{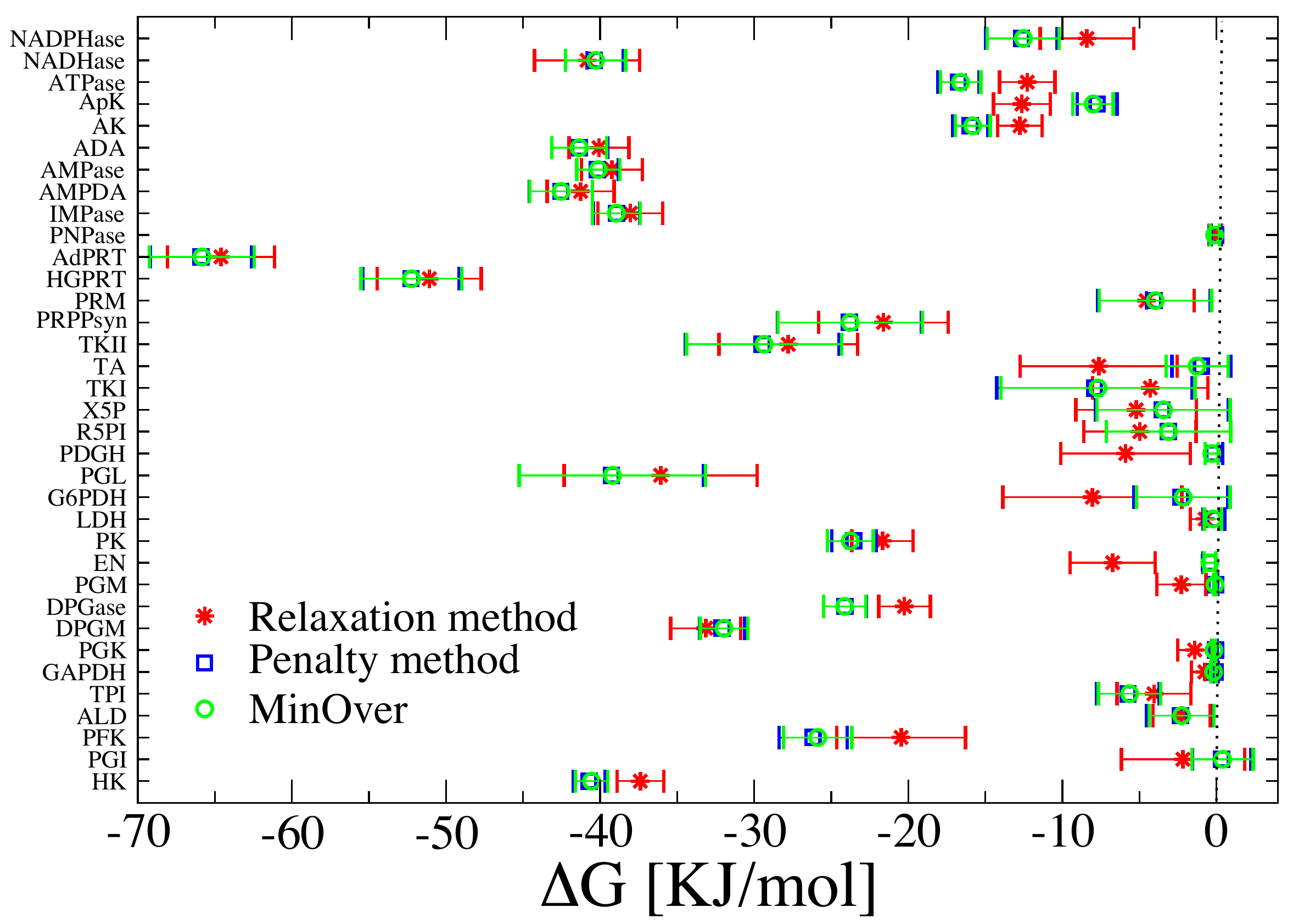}
\caption{{\bf Estimated Gibbs energy changes of reactions in the hRBC metabolic network: comparison of different algorithms.} The results obtained starting from direction assignments corresponding to MBE solutions are shown here for three different methods: (a) MinOver with $\lambda=0.01$ (this paper, free markers); (b) a penalty method using the Euclidean distance between the solution and the prior as the cost function (blue markers); (c) a relaxation method optimized to be faster (red markers). For (b), we have set $\lambda_1=0.001$ while $\lambda_2$ is initialized at $10$ and grows in steps of $10$ each time a minimum is found, until the configuration is feasible. These results show that MinOver roughly minimizes the average Euclidean distance between the solution and the prior (as the penalty method), albeit with running times over $100$ times shorter (see text for details).}
\end{center}
\end{figure}
In short, algorithm (b) minimizes the Euclidean distance between the prior $\boldsymbol{\mu}(0)$ and the solution under the thermodynamic constraint, which is enforced through the term proportional to $\lambda_2$. The relaxation method (c) simply corresponds to a particular choice of the constant $\lambda$ appearing in (\ref{deltag}): in specific, the step size is required to be proportional to the amount by which the least satisfied constraint is violated. One sees that the penalty method and MinOver produce almost identical solutions. Measuring explicitly the average Euclidean distance between the prior and the solution (the average being taken over the choice of the priors), one indeed finds $\overline{d}\simeq 15\mathrm{~KJ/mol}$ (penalty method) versus $\overline{d}\simeq 15.2\mathrm{~KJ/mol}$ (MinOver), while $\overline{d}\simeq 22.4\mathrm{~KJ/mol}$ for the relaxation method (c). Note that the gap between the performance of MinOver and that of the penalty method ($0.2\mathrm{~KJ/mol}$, corresponding to a relative error on $\overline{d}$ of just over 1\%) provides a very rough estimate of the average distance between the solutions obtained by MinOver and those obtained by cost function minimization, and is much smaller than the spread of the initial configurations of chemical potentials, which in this case is about $15\mathrm{~KJ/mol}$. Comparing running times for this case, moreover, one sees that MinOver is about $140$ times faster than the penalty method ($19$ versus roughly $2700$ seconds), while the standard relaxation method is even faster (about $2.4$ seconds).

\subsection*{Identifying and removing thermodynamically infeasible loops in the {\it Escherichia coli} metabolic network model iAF1260}

The application that we have just discussed shows that the algorithm we present can provide information on the Gibbs energy landscape, even correcting inconsistent input knowledge. We shall now employ the procedure outlined in the Materials and Methods section to efficiently identify and eliminate loops from thermodynamically infeasible flux configurations of the reconstructed metabolic network of {\it Escherichia coli} iAF1260 \cite{bigcoli}. We shall focus specifically on the periplasmic and cytoplasmic core of the network, which presents the advantage that the cycles identified here are independent of the transport and environment selections. The network includes $1767$ reactions among $1349$ chemical species.

Since we are not focusing on the reconstruction of the Gibbs energy landscape but simply on the existence of solutions of (\ref{constr}), a detailed biochemical prior is not needed. Therefore, for the present purposes we have taken $P_0(\boldsymbol{\mu})$ to be  product of identical uniform distributions. To begin with, we have fixed the direction of reactions that are putatively irreversible in the reconstructed network ($1475$ in total) and verified that this assignment is indeed thermodynamically feasible by finding a solution of (\ref{constr}) restricted to irreversible processes. Then we  integrated the above assignments by fixing randomly and independently with equal probability the directions of the $292$ processes that are putatively reversible. A large ensemble of  $\mathbf{u}$ vectors ($10^5$ instances) thus obtained was tested for thermodynamic feasibility. Note that by excluding the possibility that reactions are not operating we are considering a worst-case scenario in which all reactions bear a non-zero flux. Only about 1.5\% of these configurations turned out to be thermodynamically feasible, i.e. free from loops. We focus on infeasible instances, for which no vector $\boldsymbol{\mu}$ of chemical potentials was found to satisfy (\ref{constr}), applying to them the loop identification and removal protocol. Figure 5 shows the histogram of convergence times for the above procedure, i.e. the times required to verify that a given vector of flux directions is thermodynamically infeasible and to correct it. 
\begin{figure}
\begin{center}
\includegraphics[width=8.5cm]{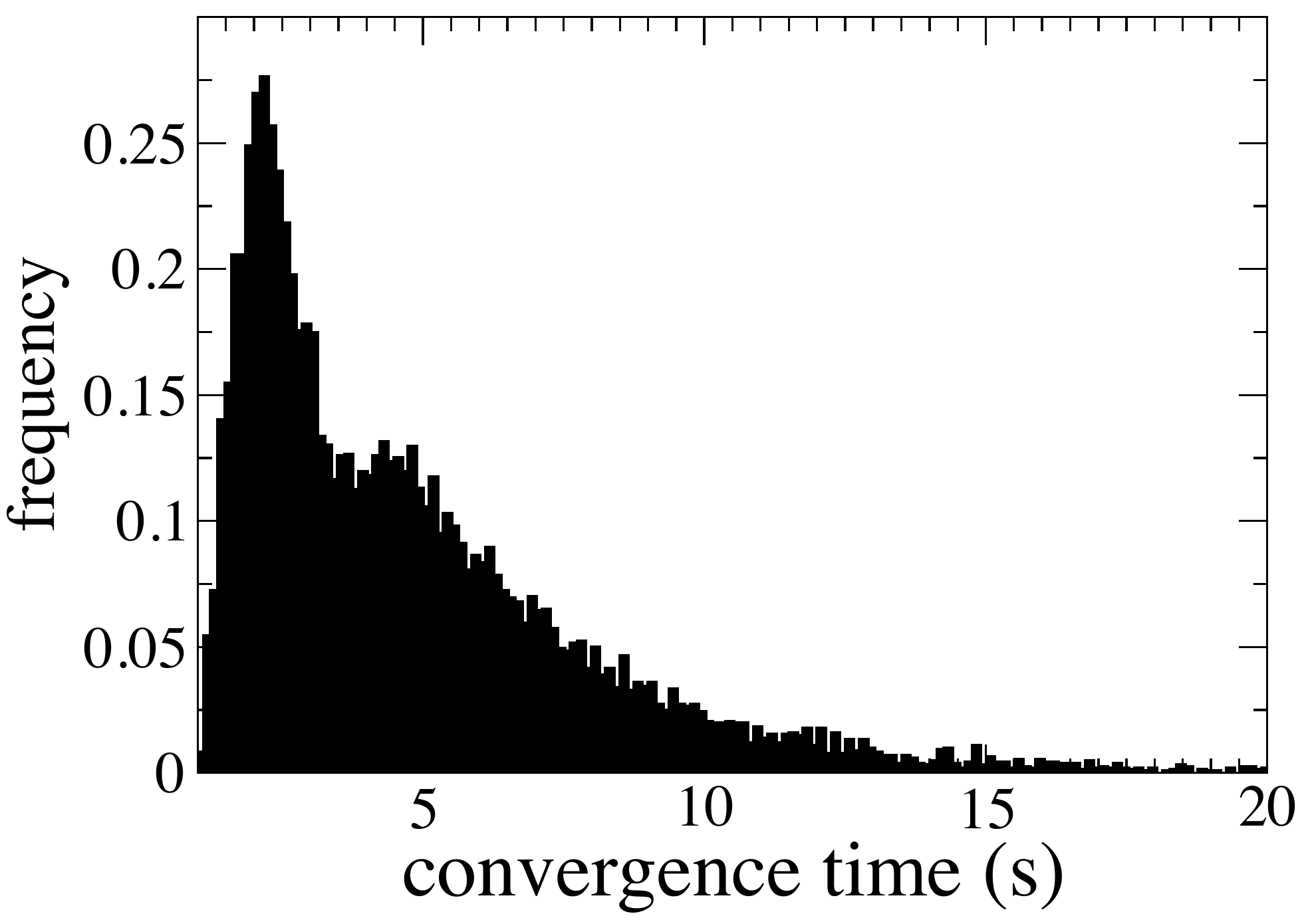}
\caption{{\bf Histogram of the convergence times of the algorithm.} Convergence times shown are for the identification and elimination of the thermodynamically infeasible loops and for the verification of thermodynamic feasibility of randomly generated flux configurations from the {\it Escherichia coli} iA1260 metabolic network model (on an Intel dual core at 3.06 GHz).}
\end{center}
\end{figure}
On an Intel dual core at 3.06 GHz the average  CPU time for convergence is of the order of a few seconds, while it exceeds 10 seconds in about 5\% of the (random) instances. In the worst case within our ensemble, convergence time was around 100 seconds. 

We have furthermore studied the set of loops that were thus identified and corrected. Quite remarkably  this analysis revealed that thermodynamical infeasibility  is related to the presence of  a small set of cycles, 23 in total. These are reported in Table \ref{cyclelist} and three of them are depicted explicitly in Figure 6. 
\begin{figure}
\begin{center}
\includegraphics[width=8.5cm]{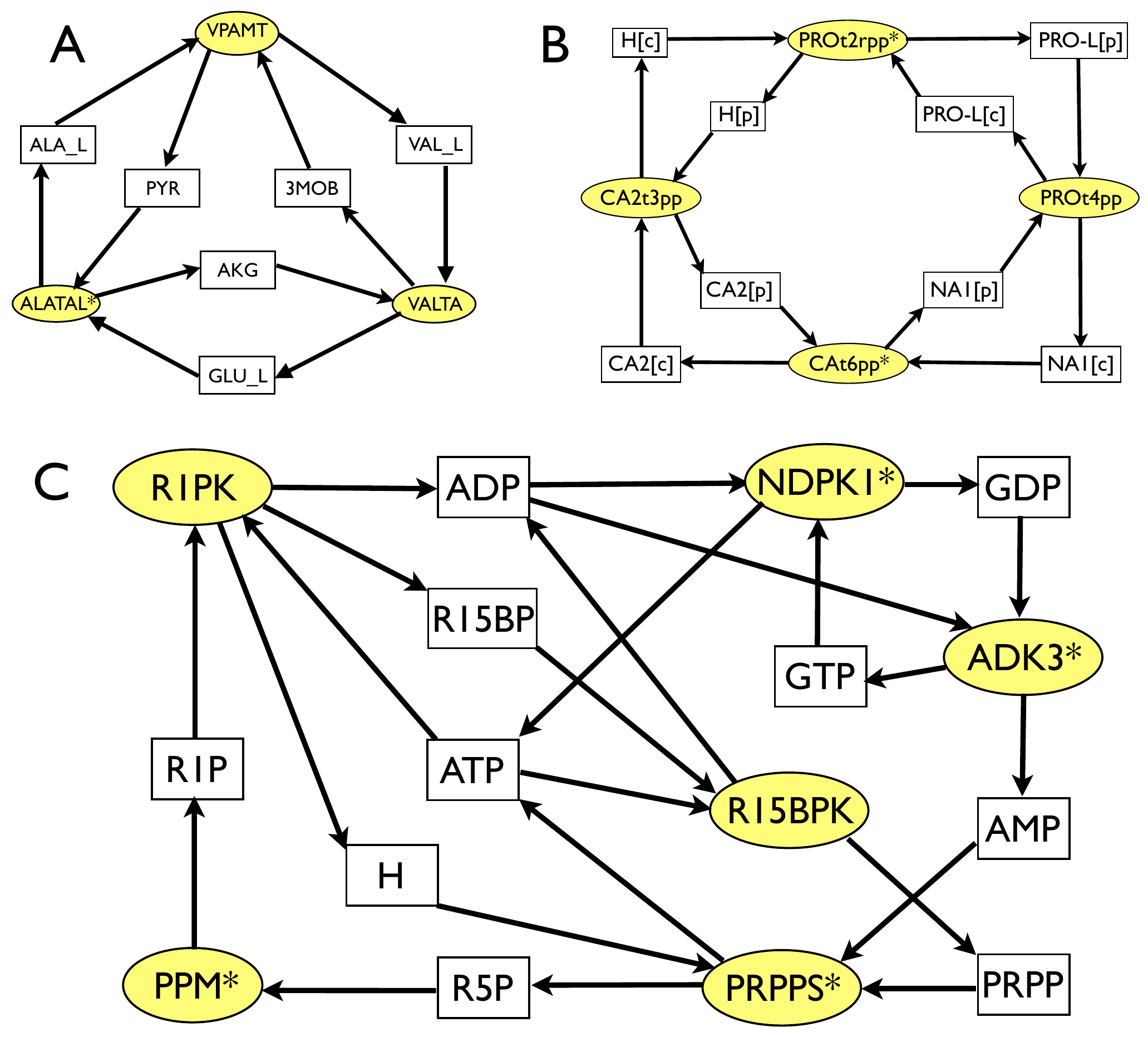}
\caption{{\bf Three of the 23 thermodynamically infeasible cycles identified in the \textit{E. coli} metabolic network iAF1260.} Rectangles (resp. ellipses) denote metabolites (resp. reactions). The cycles depicted here are n. 8 (A, top left), 18 (B, top right) and 22 (C, bottom) from Table \ref{cyclelist}. The star indicates reversible reactions according to \cite{bigcoli}. See Supporting Table S1 for abbreviations.}
\end{center}
\end{figure}
Note that some of these cycles include a single reversible reaction. This implies that in order to ensure that such cycles will not be present in the final configuration it is necessary to fix the direction of the reactions SERt2rpp, GLUt2rpp, ACt2rpp, GLYCLTt2rpp, THRt2rpp, SUCOAS, PPAKr and PROt2rpp opposite to that shown in Table \ref{cyclelist} (see the Supporting Table S1 for abbreviations). In turn, this is easily seen to impose a further constraint on the direction of the reversible fluxes CAt6pp and GLUABUTt7pp (via cycles of length 4). Finally we checked that excluding these loops guarantees thermodynamic feasibility of $10^6$ randomly generated flux configurations

\section*{Discussion} 

Ideally, constraint-based models of metabolic activity allow to appraise the energetic potential of cells based on minimal constraints related to local mass-balance and thermodynamic feasibility rules, possibly complemented with optimization principles that can encode for functional constraints. As a result, the flow of matter in non-equilibrium steady states could be characterized in terms of the Gibbs energy change of reactions, which specifies the directionality of interconversions, and of the average number of turnovers per time per volume, i.e. the flux, without the need of detailed information on enzyme kinetics or transport mechanisms. Thermodynamic constraints, strongly linked to overall intracellular conditions like ionic strength and pH \cite{stock}, are particularly subtle and rich of consequences. It has indeed been argued that the Gibbs energy landscape contains important regulatory information \cite{kummel2}. Reactions far from equilibrium are expected to be roughly insensitive to fluctuations in metabolite concentrations, so that they will be driven mostly by enzyme regulation. On the other hand, reactions close to equilibrium (i.e. with a net Gibbs energy change close to zero) bear a high sensitivity to variations in metabolite levels and are therefore unlikely targets for tight regulation. Besides, knowledge of reaction free energies (or more precisely of the chemical potentials of the metabolites involved) provides clues on metabolite levels which would be hard to obtain from mass-balance constraints only. Therefore, developing effective procedures to deal with the complexity of flux models encompassing both mass and energy constraints at genome scale is a central challenge for computational systems biology. 

Many important steps have been taken recently to tackle it. At one level, thermodynamic feasibility can be translated into topologic constraints (`absence of loops') for the flux configuration emerging from mass-balance constraints \cite{expa}. This suggests than an improvement in reversibility assignments (e.g. along the lines of \cite{kummel}) can be a key to ensure energy balance a priori in metabolic network reconstructions, with the caveat that the possibility that a reaction reverses can depend on the boundary conditions (e.g. the external supply of a certain metabolite) or on intracellular perturbations (e.g. a knockout causing the accumulation of an intermediate metabolite) \cite{hoppe}. Another possibility consists in building mixed integer-linear constraint-based models that include thermodynamic requirements in the form of consensus rules (using information on standard Gibbs free energies) \cite{tbmfa} or as additional constraints on metabolite levels (using information on measured intracellular concentrations) \cite{hoppe}. Here we have followed a different though related route that in our view complements the approaches just described. The starting point is the fact that, given a flux configuration, thermodynamic constraints can be written as simple stoichiometric inequalities for the chemical potentials. Feasibility implies the existence of a vector of chemical potentials satisfying such system. We have presented an algorithm that is able to construct solutions starting from a possibly limited and noisy biochemical prior. Our approach differs substantially from previous methods in that it relies on modifying the structure of correlations between chemical potentials (after fixing some variables to set a scale) using the stoichiometric information on reaction connectivity to drive the updating process. In this sense, the important difference with \cite{kummel2} is that the prior information is used only to initialize the algorithm, and a large flexibility is allowed in deforming it until a viable solution is obtained.

The usefulness of the algorithm in the analysis of genome-scale networks has been tested in two different cases. For the metabolic network of the human red blood cell, our approach has proved capable of reconstructing the Gibbs energy landscape correcting inconsistent prior information. In turn, this has lead to predictions for intracellular metabolite levels. It is important to stress that the bounds on concentrations we have obtained (which vary rather heterogeneously across compounds) only reflect stoichiometric information. For the metabolic network of {\it Escherichia coli}, instead, we have focused on the problem of correcting thermodynamically infeasible flux states in the core formed by the periplasmic and cytoplasmic matrix. We have thoroughly analyzed a large ensemble of configurations of reaction directions, identifying the cycles responsible for thermodynamic inconsistency and correcting them in a very modest amount of CPU time. Quite intriguingly, we have related infeasibility to the existence of a relatively small number of short cycles in the flux configuration, whose removal suffices to ensure thermodynamic feasibility in worst-case flux configurations.

The main advantage of our method consists in our view in its efficient implementation. On the critical side, we point to two aspects that deserve further study. In first place, our tool requires flux configurations as inputs, i.e. it is still unable to produce thermodynamically feasible configurations of fluxes and chemical potentials starting from no previous reversibility hypothesis. However it may provide the basis of a more general procedure for the analysis of genome-scale metabolic networks that couples flux- and thermodynamic profiling, a challenging open problem in computational biology.  Secondly, our method relies on prior biochemical information and it would be desirable for it to be effective even if much or most biochemical priors are unknown. As we pointed out, some information has to be injected into the problem for the sake of definiteness. The interesting question is therefore what is the minimum necessary prior needed to reconstruct the Gibbs energy landscape and how are predictions affected by restricted priors. Such problems are mathematical in nature and could bear a particularly high significance for modeling purposes. 

We remark that the corrected flux configurations thus obtained, like the starting ones (which were drawn from a uniform product measure over reactions), are not guaranteed to be consistent with any steady state assumption. On the other hand, see Supporting Text S1, starting from a mass-balanced configuration one retrieves another mass-balanced configuration. Clearly, a method that directly generates thermodynamically steady state flux vectors and chemical potentials would be highly welcome. Nevertheless, we note that our analysis focuses on a worst-case scenario where all reactions bear a non-zero flux. In a more realistic case where reactions may be inactive it is reasonable to expect that the flux directions we generate will allow for a steady state. In turn, the identification of these loops might be weakly conditioned on the sample of flux assignments we employed. We can't rule out the possibility that assigning directions based e.g. on mass balance constraints provides a selection criterion for flux configurations that differs substantially from the uniform measure we employed. However we expect that our sample allows to correctly identify loops involving at most $\log_2 S\simeq 16$ reversible reactions, $S$ being the number of distinct direction assignments in our sample. Only loops that include a larger number of reversible processes might therefore play a role in a differently selected set of direction assignments.    

~

{\bf Acknowledgments --} This work was supported by the DREAM Seed Project of the Italian Institute of Technology (IIT) and by the joint IIT/Sapienza Lab ÒNanomedicineÓ. The IIT Platform ``Computation'' is gratefully acknowledged.

\begin{table*}
\begin{center}
\begin{tabular}{| c | c | c | }  
\hline 
\bf{Cycle ID} & \bf{Lenght} & \bf{Formula}  \\ 
\hline 
1&3  &  SERt4pp  $+$     NAt3pp  $-$   SERt2rpp(R)   \\
2&3  &  NAt3pp  $+$     GLUt4pp  $-$     GLUt2rpp(R)   \\
3&3  &  NAt3pp  $-$ ACt2rpp(R)   $+$     ACt4pp  \\
4&3  &  NAt3pp  $-$     GLYCLTt2rpp(R)  $+$     GLYCLTt4pp   \\
5&3  &  PROt4pp $-$   PROt2rpp(R)  $+$      NAt3pp   \\
6&3  &  HPYRRx $-$ TRSARr(R)  $-$     HPYRI(R)     \\
7&3  &  CRNDt2rpp(R)  $-$ CRNt2rpp(R)  $+$     CRNt8pp    \\
8&3  &  VPAMT  $-$     ALATAL(R)  $+$     VALTA(R)  \\
9&3  &  ABUTt2pp  $+$     GLUABUTt7pp(R)  $-$     GLUt2rpp(R)  \\
10&3 & NAt3pp$-$ THRt2rpp(R) $+$ THRt4pp\\
11&3 & ADK3(R) $-$ ADK1(R) $+$ NDPK1(R) \\
12&3 & ACCOAL $+$ PPCSCT $-$ SUCOAS(R) \\
13&3 & PPAKr(R) $+$ ACCOAL $+$ PTA2 \\
14&4  &  ACt4pp  $-$     CAt6pp(R)  $-$     ACt2rpp(R)  $+$     CA2t3pp  \\
15&4  &  CA2t3pp  $-$   GLYCLTt2rpp(R)  $+$  GLYCLTt4pp  $-$     CAt6pp(R)  \\
16&4  &  SERt4pp     $-$    CAt6pp(R)     $-$    SERt2rpp(R)     $+$    CA2t3pp\\
17&4  &  GLUt4pp     $-$    CAt6pp(R)     $+$    CA2t3pp  $-$    GLUt2rpp(R)  \\
18&4  &   CA2t3pp     $-$    PROt2rpp(R)     $+$    PROt4pp     $-$    CAt6pp(R)    \\
19&4  &   THRt4pp $-$ CAt6pp(R) $-$ THRt2rpp(R) $+$ CA2t3pp   \\
20&5  &   ADK1(R) $-$ ACKr(R) $+$ ACS $+$ PTAr(R) $-$ PPKr(R)   \\
21&5  &   R15BPK $-$ PPM(R) $-$ PRPPS(R) $-$ ADK1(R) $+$ R1PK  \\
22&6  &  R1PK  $-$ NDPK1(R)  $-$  PPM(R)  $-$     PRPPS(R)  $+$     R15BPK  $-$     ADK3(R)  \\
23&6  &  ADK3(R) $-$ ACKr(R) $+$ PTAr(R) $+$ NDPK1(R) $-$ PPKr(R) $+$ ACS  \\
\hline
\end{tabular}
\caption{{\bf Thermodynamically infeasible cycles for the \textit{E. coli} metabolic reaction network iAF1260.} The list provides the complete set of cycles turned out in the thermodynamic feasibility analysis of a sample of $10^5$ different randomly generated flux configurations. Plus (resp. minus) signs indicate that the reaction participates in the cycle in its forward (resp. backward) direction. (R) indicates that the corresponding reaction is putatively reversible according to \cite{bigcoli}. Analyzing interdependencies in the above cycles, one sees that the directions of the putatively reversible fluxes SERt2rpp, GLUt2rpp, ACt2rpp, GLYCLTt2rpp, THRt2rpp, SUCOAS, PPAKr, PROt2rpp, CAt6pp and GLUABUTt7pp are in fact constrained. See Supporting Table S1 for abbreviations.}
\label{cyclelist}
\end{center}
\end{table*}

\end{document}